\documentclass[12pt]{elsarticle}

\RequirePackage{fix-cm}

\makeatletter
\let\@afterindenttrue\@afterindentfalse
\makeatother

\newif\ifdraft
\drafttrue

\makeatletter
\def\ps@pprintTitle{
 \let\@oddhead\@empty
 \let\@evenhead\@empty
 \def\@oddfoot{\centerline{\thepage}}
 \let\@evenfoot\@oddfoot}
\makeatother

\usepackage{graphicx}
\usepackage{svg}
\usepackage{tablefootnote}

\RequirePackage{tikz}

\usepackage{diagbox}

\usetikzlibrary{shapes,arrows,positioning,calc,fit,intersections,through,graphs, bending}

\definecolor{sfcol}{HTML}{80b1d3}

\tikzset{
  grammar/.style={
    draw, rectangle, minimum width=18mm, minimum height=6mm, align=center
  },
  mutations/.style={
    draw, dotted, rectangle, minimum width=18mm, minimum height=6mm, align=center
  },
  shell/.style={
    draw, ellipse, minimum width=13mm, minimum height=6mm, align=center
  },
  sfcomponent/.style={
    fill=sfcol, fill opacity=0.4, text opacity=1, rectangle, align=center
  },
}

\usepackage{fontawesome}
\newcommand{\suppOK}[1][green]{\text{\color{#1}\faCheck}}
\newcommand{\suppNO}[1][red]{\text{\color{#1}\faClose}}

\usepackage{libertine}
\usepackage{hyperref}

\usepackage{amsmath,amssymb}
 
\usepackage[scaled=0.81]{beramono}
\usepackage{xparse,xspace}

\NewDocumentCommand{\Tool}{m}{\texttt{#1}\xspace}
\newcommand{\AFL}{\Tool{AFL++}}
\newcommand{\mksh}{\Tool{mksh}}

\NewDocumentCommand{\Gen}{m}{\ensuremath{#1}\xspace}
\newcommand{\genI}{\Gen{I}}
\newcommand{\genM}{\Gen{M_I}}
\newcommand{\genV}{\Gen{V}}

\newcommand{\param}[1]{\ensuremath{\textsl{#1}}}
\newcommand{\proc}[1]{\ensuremath{\textsf{#1}}}
\newcommand{\PROCEDURE}[3]{\STATE \textbf{procedure} #1 (#2) $\to$ #3}

\usepackage{relsize}
\NewDocumentCommand{\shf}{s}{\IfBooleanTF{#1}{ShellFuzzer}{{\smaller[0.5]\textsc{ShellFuzzer}}}\xspace}

\usepackage{fontawesome}

\usepackage{color} 

\definecolor{redcol}{RGB}{215,25,28}
\definecolor{orangecol}{RGB}{253,174,97}
\definecolor{lightbluecol}{RGB}{171,217,233}
\definecolor{darkbluecol}{RGB}{44,123,182}
\colorlet{addcolor}{redcol}

\colorlet{aflcol}{redcol}
\colorlet{kleecol}{darkbluecol}
\colorlet{cbmccol}{orangecol}

\usepackage{numprint}

\usepackage[inline]{enumitem}
\setlist[enumerate]{label=\emph{\roman*})}
\setlist[description]{font=\normalfont\bfseries}

\usepackage{amsmath}

\usepackage{framed}

\usepackage{hanging}

\usepackage[T1]{fontenc}
\usepackage{float}

\usepackage{listings}
\usepackage{tcolorbox}
\tcbuselibrary{listings}
\newcommand{\lcolorbox}[2]{
  \hspace*{-\fboxsep}\colorbox{#1}{#2}
}
    
\newcommand{\highlighcode}[1]{\lcolorbox{lightbluecol!70}{#1}}

\newcommand{\Sh}[1]{\mbox{\lstinline[basicstyle=\ttfamily,language=bash]|#1|}}
    
\usepackage{pgfplotstable,filecontents}
\pgfplotsset{compat=1.9}
\usepackage{array}

\usepackage{graphicx} 
\usepackage{booktabs,multirow} 
\usepackage{hyperref} 
\usepackage{tabularx}

\usepackage{csvsimple}
\usepackage{longtable}
\usepackage{booktabs}
\usepackage{caption}
\usepackage{subcaption}
\usepackage{backnaur}
\usepackage{tcolorbox}

\newtcolorbox{tbox}[1]{colback=white, colframe=black, fonttitle=\bfseries, title=#1}
\usepackage{hyphenat}
\usepackage{algorithmic}

\DeclareDocumentCommand{\finding}{m}{
	\begin{tcolorbox}[
		boxsep=0pt,
		left=6pt,
		right=6pt,
		top=2pt,
		bottom=2pt,
		colframe=white,
		colback=lightbluecol!70!white,
		boxrule=0.3pt, 
		]
		\small\textit{
			#1
		}
	\end{tcolorbox}
}

\begin{document}

\begin{frontmatter}



\title{ShellFuzzer: Grammar-based Fuzzing\\ of Shell Interpreters}



\author[inst1]{Riccardo Felici}

\affiliation[inst1]{organization={Faculty of Informatics, USI Università della Svizzera italiana},
            city={Lugano},
            country={Switzerland}}

\author[inst1]{Laura Pozzi}
\author[inst1]{Carlo A. Furia}


\begin{abstract}

  Despite its long-standing popularity and fundamental role in an operating system, the Unix shell has rarely been a subject of academic research.
  In particular, regardless of the significant progress in compiler testing, there has been hardly any work applying automated testing techniques to detect faults and vulnerabilities in shell interpreters.

  To address this important shortcoming, we present \shf: a technique to test Unix shell interpreters by automatically generating a large number of shell scripts.  \shf combines grammar-based generation with selected random mutations, so as to produce a diverse range of shell programs with predictable characteristics (e.g., valid according to the language standard, and free from destructive behavior).
  
  In our experimental evaluation, \shf generated shell programs that exposed 8 previously unknown issues that affected a recent version of the \mksh POSIX-compliant shell; the shell maintainers confirmed 7 of these issues, and addressed them in the latest revisions of the shell's open-source implementation.
\end{abstract}


\begin{keyword}
Shell interpreters \sep Testing \sep Fuzzing \sep Grammar-based 
\end{keyword}

\end{frontmatter}

\section{Introduction}

With all its shortcomings, quirks, and obsolete features, the (Unix) shell remains a ubiquitous tool
to program automation and administrative tasks. The shell's popularity and its role as a critical component underlying so many different computer systems entail that its security and reliability are paramount and, conversely, any vulnerability\footnote{\url{https://www.cve.org/CVERecord?id=CVE-2014-6271}} may have a disproportionate harmful impact~\cite{ss-study}.  Given the situation, it is somewhat surprising that the shell has been largely neglected as a research topic~\cite{GreenbergKV21}.

Motivated by the shell's practical importance, in this paper we present \shf: a technique to \emph{test} shell interpreters by automatically generating shell scripts with different characteristics.
Similarly to what is done to test compilers of other languages (see \autoref{sec:testing-techniques}), \shf generates shell programs guided by a \emph{grammar} of the shell language. In order to increase the diversity of the produced shell scripts, \shf combines three separate \emph{generators}, each using a different fragment of the POSIX shell syntax, and may also apply random mutations that introduce small syntactic errors into the produced scripts.

Combining three generators with different characteristics also helps \shf produce \emph{oracles} that capture some fundamental behavioral properties of each generated scripts.  Precisely, one generator always produces valid shell scripts, whereas the two other generators always produce invalid shell scripts; correspondingly, the former's oracles assert that execution terminates without errors or abrupt termination, whereas the latter's require that the shell interpreter detects the invalid code and signals it with a suitable error message or return code.
In addition to these generator-specific oracles, \shf can also detect generic memory errors by deploying standard memory sanitizers during the testing process.

Since a shell interpreter usually has full access to the environment where the tests are run, an automatically generated shell script may accidentally disrupt the testing process by executing destructive operations (such as deleting large chunks of the file system).
In order to minimize the chance of this happening, we carefully designed \shf's generators so that they omit certain commands and features that are especially likely to introduce destructive behavior.
While this somewhat restricts the range of programs that \shf generates, our experiments suggest that it is a practical trade off because it retains a good effectiveness of testing while also avoiding time-consuming destructive crashes.

We implemented \shf in a tool with the same name,
and we evaluated its capabilities with experiments targeting open-source shell interpreters.  \shf managed to find 8 unique bugs in the \mksh shell, 7 of which were confirmed by the shell maintainers (including 2 memory errors that also affected Bash).  Our experiments also demonstrate that \shf's generators are complementary, enabling it to detect different kinds of bugs.

\paragraph{Contributions}
This article presents the following main contributions:

\begin{itemize}
\item \shf, an automated testing technique for POSIX shell interpreters.

\item A prototype implementation of the \shf technique in a tool with the same name, which is publicly available.\footnote{\url{https://github.com/user09021250/shellfuzzer/}}

\item An empirical evaluation of \shf, demonstrating its capabilities of finding real bugs in shell interpreters, and how they compare to using general-purpose fuzzing techniques.
\end{itemize}

\paragraph{Organization}
The rest of the paper is structured as follows.
\autoref{sec:related} summarizes the main approaches to interpreter/compiler testing, and the (little) related work targeting the shell.
\autoref{sec:methodology-implementation} presents the design of \shf, motivated by the key challenges of effectively testing shell interpreters.
\autoref{sec:exp-design}
introduces the research questions that capture \shf's practical capabilities, and outlines the experiments we ran to answer those questions.
\autoref{sec:exp-results}
presents the results of experiments, and summarizes the key observations about \shf's capabilities.
\autoref{sec:conclusion} concludes with a high-level summary of the paper's contributions, and an outline of future work.

\section{Related Work}
\label{sec:related}

We summarize the main related work in two areas: shell languages and programming (\autoref{sec:shell_interpreters}), and the application of testing and analysis techniques to language interpreters and compilers (\autoref{sec:testing-techniques}).

\subsection{Shell Languages and Interpreters}
\label{sec:shell_interpreters}

A shell is a command-line interpreter that allows users to access the services of operating systems.  This paper focuses on Unix shells, which have been historically prominent and remain widely available and used.  The first Unix shell was developed over fifty years ago, as part of the Unix operating system~\cite{shell2}.  As Unix gained broader popularity in the late 1970s, the Bourne shell (\Sh{sh})\footnote{\url{https://www.in-ulm.de/~mascheck/bourne/}} became the standard shell, followed by Bash\footnote{\url{https://www.gnu.org/software/bash/}} some ten years later -- which quickly became the standard shell on Linux, where it remains ubiquitous.  Other Unix shells of historical or practical importance include the Almquist shell (\Sh{ash}),\footnote{\url{https://www.in-ulm.de/~mascheck/various/ash/}} the KornShell (\Sh{ksh}),\footnote{\url{http://kornshell.com/}} the Z shell (\Sh{zsh}),\footnote{\url{https://www.zsh.org/}} and the C shell (\Sh{csh}).\footnote{ \url{https://github.com/freebsd/freebsd-src/tree/main/bin/csh} }
The POSIX standard~\cite{posix} includes the specification of a fundamental set of features that most shells are (at least partially) compliant with -- even though each shell typically provides its own custom variants and extensions on top of a POSIX core.

Despite its longevity and ubiquity -- Bash even is in the top-50 of most popular programming languages\footnote{\url{https://www.tiobe.com/tiobe-index/}} -- the shell ``has been largely ignored by academia and industry''~\cite{GreenbergKV21}:\footnote{As a result, several of the references in this section are URLs instead of refereed publications.} there is little research work that targeted the rigorous analysis of shell scripts, even though it is generally known that writing shell scripts is error prone~\cite{bash_bugs}.\footnote{\url{https://mywiki.wooledge.org/BashPitfalls}}
Among the few exceptions, ABash~\cite{ABash} is a static analyzer for Bash scripts based on abstract interpretation; even though it does not provide soundness or completeness guarantees (due to the highly dynamic nature of the shell language), it was effective at finding bugs in shell scripts used in various open-source projects.  CoLiS~\cite{CoLiS} is a formally verified interpreter for a core shell language, which has been used to find numerous bugs in Debian maintainer scripts~\cite{CoLiS-2}.  Smoosh~\cite{shell_semantics} is a formal, mechanized, executable semantics of the POSIX shell language, which can be used to rigorously evaluate POSIX conformance, as well as to discover conformance bugs in shell interpreters.  NoFAQ~\cite{NoFAQ} is a synthesis-based automated repair tools for command line interfaces (not specific to shell languages).  Practitioners have also released several tools that help users write correct and maintainable shell scripts.  For example, Modernish\footnote{\url{https://github.com/modernish/modernish}} is a library that provides safer, more convenient variants of some shell language features.  ShellCheck\footnote{\url{https://www.shellcheck.net/}} is a linter for shell scripts.  ExplainShell\footnote{\url{https://explainshell.com/}} is a documentation system that can link the relevant information about each part of a shell command.

There has also been research on improving the flexibility -- e.g., \cite{shell_perf} -- and performance -- e.g., \cite{LiargkovasK0V23,shell_perf2,bash_accel} -- of shells.  Since such research is less relevant for the present paper we do not go into the details of these works, but we refer the interested readers to the recent discussion at HotOS~\cite{shell-future,GreenbergKV21}.

\subsection{Testing Interpreters and Compilers} 
\label{sec:testing-techniques}

Chen et al.'s recent survey of compiler testing~\cite{compiler_testing_survey2}
highlights several key challenges that also apply to interpreter testing, and provides a general motivation for our approach to shell interpreter testing.  First of all, being able to reliably generate \emph{valid} programs is a key challenge.  To this end, the vast majority of approaches (e.g., \cite{csmith,sound-opt,lang_fuzz}) are grammar-based, that is they use a formal grammar of (a fragment of) the input language to generate programs that are (syntactically) valid.

Using a grammar to drive the generation of programs helps generate consistently valid programs; however, it may prevent a more effective exploration of the program space.  Besides, sometimes being able to add a modicum of invalid, ambiguous, or undefined features is useful to increase the diversity of the generated programs, and to test the interpreter or compiler more thoroughly.  To this end, several more recent approaches (e.g., \cite{grammar_mutation_strategies,LeSS15-b,SunLS16,ChenSSSZ16,DonaldsonELT17,superion}) combine grammar-based generation with some form of \emph{mutation}.  Our \shf also uses a combination of grammar-based generation and mutations in order to increase the diversity of shell scripts it can generate while retaining control on their validity and features.

Even though grammar-based approaches
are widespread for interpreter/compiler testing (especially in recent times),
other testing techniques have also been successfully applied to these challenges~\cite{EideR08,PalkaCRH11,NagaiHI14}.  In particular, combinations of random testing/fuzzing~\cite{fuzzing_another_survey,surveyimp,fuzzing_recent_survey} with exploration heuristics remain widely used as they can still be highly effective~\cite{compiler-fuzzing,yarpgen1,ChenGZWFER13}.  For this reason, in \autoref{sec:results:rq3} we will experimentally compare our \shf tool to the modern, popular fuzzing framework \AFL.  In related efforts, fuzzing~\cite{fuzzing-unix-utils} and other test-input techniques such as symbolic execution~\cite{klee} have been routinely used to test the reliability of operating-system command-line utilities or components~\cite{semantic_bugs_in_fs}.

Another key challenge of any kind of testing is generating oracles.  Implicit oracles~\cite{oracle-survey} -- such as crashing oracles, or memory safety checkers~\cite{memorysanitizer} -- are commonplace since they are always available.  However, the correct behavior of an interpreter/compiler involves much more than just avoiding crashes or memory errors.  Then, differential and metamorphic testing are commonly used techniques to automatically derive more expressive oracles in interpreter/compiler testing.  In \emph{differential} testing (e.g., \cite{TaoWZS10,LeSS15,ChenSSSZ16,sound-opt,SunLS16-b}), one runs the same generated programs with multiple compilers (or the same compiler with different optimization options), and reports any inconsistency in how the compiled binaries execute.  In \emph{metamorphic} testing (e.g., \cite{DonaldsonELT17,orange4,orion_equivalence_modulo_input}), one runs equivalent variants of the same program (for example, a program with and without dead code) and checks whether they behave in the same way.  As we'll explain in \autoref{sec:methodology-implementation}, \shf uses a combination of implicit oracles (to detect basic memory errors) and automatically generated oracles linked to the characteristics of each generated shell program.

In terms of targeted \emph{languages}, the vast majority of research on compiler/interpreter testing targets statically typed languages.  Out of the 42 approaches listed in the survey~\cite{compiler_testing_survey2}'s Table~2, only 4 target the well-known dynamic languages JavaScript~\cite{lang_fuzz,l2f,GroceAZCR16,Bastani0AL17}, Python and Ruby~\cite{Bastani0AL17}.  In the last few years, there has been more work on testing JavaScript engines~\cite{superion,js_seed_selection} and Python runtimes~\cite{python_interpreter_fuzzing} -- often using a tried-and-true combination of grammar-based generation and mutations.  However, hardly any research work applied automated testing techniques to shell interpreters: \autoref{sec:shell_interpreters} mentions the only exception we know of~\cite{shell_semantics}, which is based on a detailed mechanized semantics of the POSIX shell language.  In this work, we want to apply grammar-based fuzzing techniques with the goal of detecting not only low-level memory and crashing bugs, but also logic bugs affecting specific functionality of a shell interpreter's implementation.  Doing so involves several of the same challenges of compiler testing, while dealing with a highly-dynamic, idiosyncratic language.  An additional challenge, specific to shells, is that a shell script can easily modify any parts of the environment that executes the test.  In the next section, we explain how \shf addresses these challenges.

\section{Design and Implementation of \texorpdfstring{\shf*}{ShellFuzzer}}
\label{sec:methodology-implementation}

\shf produces shell scripts using a grammar-based generator to test shell interpreters.  As we discussed in \autoref{sec:related}, grammar-based generation is customary to test any programs that require complex, structured inputs -- such as compilers and interpreters.  In these scenarios, a grammar helps ensure that well-formed inputs are produced, so that the testing process can exercise the program under test more extensively (beyond just crashing it with an invalid input).

A fundamental challenge of grammar-based test generation -- really, of all test generation techniques -- is the so-called \emph{oracle} problem:
after generating a shell script, and running it with a shell interpreter, how do we determine whether the test passes (behaved as expected) or fails (revealed a potential error)? One common approach is using an \emph{implicit} oracle, that is a condition that is generically applicable to every execution.  For example, we could classify a test execution as a failure whenever it ends with a crash (the so-called crashing oracle), with a non-zero exit code, or prints some error message to standard error.

A challenge of using implicit oracles together with grammar-based generation is that an oracle's validity depends on the characteristics of the input program that is generated.  For instance, if a syntactically incorrect program is rejected by an interpreter with a ``syntax error'' message, we should consider the test passing, as this is exactly what it should do; in contrast, if a syntactically correct program is rejected by an interpreter with the same error message, we should consider the test failing, as the interpreter should be able to process programs that comply with the language syntax specification.  Thus, there is a trade-off between how restricted the grammar used for generation is and how reliably one can associate implicit oracles with a generated program.  If the grammar is very constrained (for example because it only generates programs that are syntactically correct and free from any undefined behavior) it is easy to associate an implicit oracle (for example, termination without any error message); however, it will also be able to exercise less thoroughly the interpreter under test.  In contrast, if the grammar is very flexible (for example, it generates all sorts of valid and invalid programs) it is more likely to trigger unexpected behavior in the interpreter; however, automatically determining whether a generated program is valid or invalid will be impossible in general.

Another, practical challenge of grammar-based generation of shell scripts is that such scripts can easily have destructive behavior (e.g., with commands such as \Sh{rm}) since the interpreter that runs them has direct access to the system that also runs the testing infrastructure.  This determines another trade-off between generality and feasibility: generating diverse shell scripts may improve testing coverage, but it is also likely to generate destructive scripts that disrupt the testing process.  Sandboxing the execution environment guards against the worst consequences, but destructive-behavior scripts can still result in a major loss of testing time and results.

\begin{figure}[!tb]
  \centering
  \begin{tikzpicture}[node distance=10mm and 10mm]
    
    \node[grammar] (grammar) {\textsf{grammar}};
    \node[right=of grammar,mutations] (mutations) {\textsf{mutations}};
    
    \node[fit=(grammar)(mutations),sfcomponent,inner ysep=2mm,inner xsep=2mm,label={[above]\textsc{generator}}] (generator) {};
    
    \draw[-latex] (grammar) -- node[align=center] {\scriptsize{test}\\[-2pt]\scriptsize{case}} (mutations);
    \draw (mutations.east) -- (generator.east);

    \node[shell,right=of mutations] (shell) {\textsc{shell}};
    \draw[-latex] (generator.east) -- node[align=center] {\scriptsize{test}\\[-2pt]\scriptsize{case}} (shell);

    \node[right=17mm of shell,sfcomponent,label={[above]\textsc{oracle}}] (oracle) {expect:\\$\mathtt{ec} = c$, $\epsilon = e$};
    \draw[-latex] (shell) -- node[above] {\scriptsize{exit code \texttt{ec}}} node[below] {\scriptsize{error msg $\epsilon$}} (oracle);

    \node[above right=1mm and 5mm of oracle] (success) {\suppOK};
    \draw[-latex] (oracle.east) -- (success);
    \node[below right=1mm and 5mm of oracle] (fail) {\suppNO};
    \draw[-latex] (oracle.east) -- (fail);
    
\end{tikzpicture}
\caption{Overview of how each generator of \shf works.  A generator produces test cases (shell scripts) with certain characteristics, and is associated with an automatic oracle that captures the fundamental expected behavior of a shell interpreter when it executes those test cases.}
  \label{fig:shell-fuzzer}
\end{figure}
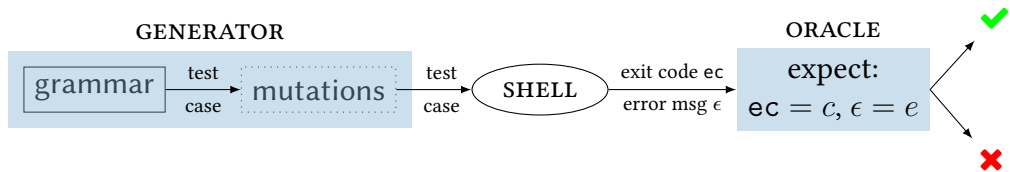

\begin{table}
  \renewcommand{\arraystretch}{1.5}
  \small
  \centering
\begin{tabular}{c *{1}{>{\raggedright\arraybackslash}p{0.19\textwidth}<{}} c *{1}{>{\raggedright\arraybackslash}p{0.16\textwidth}<{}} *{1}{>{\raggedright\arraybackslash}p{0.19\textwidth}<{}}}
  \toprule
  \multicolumn{1}{c}{\textsc{generator}}
  & \multicolumn{1}{c}{\textsc{grammar}}
  & \multicolumn{1}{c}{\textsc{mutations}}
  & \multicolumn{1}{c}{\textsc{oracle} (\textsl{expect})}
  & \multicolumn{1}{c}{\textsc{when test} \suppNO} \\
  \midrule
  \genV & POSIX-valid scripts & -- & $\mathtt{ec} = 0$, no syntax error message &
      shell rejects syntactically valid program \\
  \genI & POSIX-invalid scripts & -- & $\mathtt{ec} \neq 0$, error message &
      shell accepts invalid program \\
  \genM & POSIX-valid scripts, nested conditionals
  & syntax errors & 
  $\mathtt{ec} \neq 0$, error message &
      shell accepts invalid program \\
  \bottomrule
\end{tabular}
\caption{An overview of \shf's generators. For each \textsc{generator},
  what kinds of scripts its \textsc{grammar} generates,
  whether any \textsc{mutations} are applied, the associated automatic \textsc{oracle} (the shell interpreter behavior that we expect to observe when executing the scripts), and what it means \textsc{when} a \textsc{test} fails \suppNO\ (it is not consistent with the oracle).}
  \label{tab:generators_outcomes}
\end{table}

\subsection{How \texorpdfstring{\shf*}{ShellFuzzer} Works}
\label{sec:methodology}

The design of \shf addresses the aforementioned challenges by combining three different \emph{generators} with complementary characteristics.  Each generator uses a grammar that captures a different subset or variant of the POSIX shell standard~\cite{posix}; it may also add random mutations to the scripts generated according to the grammar.  Crucially, the grammar and mutations of each generator are chosen so that:
\begin{enumerate*}
\item they determine programs with homogeneous characteristics, which all correspond to a certain expected behavior (e.g., error message) when executed by a shell interpreter;
\item they are unlikely to generate programs with destructive behavior (e.g., they do not include destructive commands such as \Sh{rm}, or other combinations of features that are likely to lead to destructive behavior).
\end{enumerate*}
Using multiple, complementary generators gives a fine-grained control on what kinds of scripts each generator produces (thus addressing the oracle problem) while still generating scripts with diverse characteristics (thus improving the thoroughness of testing and, potentially, the kinds of bugs that it can discover).

\subsubsection{Generators \texorpdfstring{\genV, \genI, and \genM}{V, I and M\_I}}

\autoref{fig:shell-fuzzer}
illustrates how each generator is used in the testing process.  A test case is a script, first generated according to the generator's grammar, and then also possibly mutated.  Executing the script with a shell interpreter results in an exit code and in, possibly, some error message.  The generator also automatically provides an \emph{oracle} that asserts what the expected exit code is, and what error message is expected  -- depending on the characteristics of each generator's grammar and mutations.  As usual, a test is passing \suppOK\ if the oracle's assertion holds; otherwise, the test is failing \suppNO.  A failing test may be exposing an error (bug) in the shell interpreter.

Currently, \shf includes the three generators described in \autoref{tab:generators_outcomes}:
\begin{description}
\item[Generator \genV] uses a grammar that includes a significant subset
  of the POSIX standard.\footnote{
    This grammar includes:
    variable assignments, arithmetic operations and expansions (e.g. \Sh{v="$\{X\}"}, \Sh{v="$((X+Y))"}),
    builtins and other commands (e.g., \Sh{eval}, \Sh{exec}, \Sh{set}),
    function definitions,
    pipes, redirections, and other process directives.}
  This grammar should only produce scripts that are syntactically valid according to the standard; hence, when an interpreter runs a script produced by \genV, it should \emph{not} give any syntax error message (\genV's oracle).

\item[Generator \genI] uses a grammar that is based on a subset of the POSIX standard, which we manually modified to include deliberate well-formedness errors.  This grammar should only produce scripts that are \emph{invalid}  according to the standard; hence, when an interpreter runs a script produced by \genI, it \emph{should} give a syntax error message and terminate with a non-zero exit code (\genI's oracle).

\item[Generator \genM] also generates invalid scripts, but starts from a small subset of the POSIX standard grammar that defines syntactically valid programs with nested conditionals; then, it applies random mutations to the innermost statements in the program, introducing syntax errors.\footnote{ Applying mutations on a grammar of valid programs makes the kinds of invalid programs that \genM generates different from those that \genI generates.}
  Thus, \genM only produces scripts that are \emph{invalid}  according to the standard; when an interpreter runs a script produced by \genI, it \emph{should} give a syntax error message and terminate with a non-zero exit code (\genI's oracle).
\end{description}
We also curated the grammars of all generators to exclude destructive commands, and to achieve a good variety of generated programs in terms of size and used features.

\begin{figure}[!bt]
  \centering
  \begin{subfigure}[b]{0.365\linewidth}
\begin{lstlisting}[numbers=left]
for E9 in D a; do
 Ldn="$((3&4))" && echo "$Ldn";
done
exit 0;
\end{lstlisting}
    \caption{A POSIX-valid script, produced by generator \genV.}
    \label{fig:example2:V}
  \end{subfigure}
  \begin{subfigure}[b]{0.25\linewidth}
\begin{lstlisting}[numbers=left]
[ - done func87();(*\label{l2:i:error}*)
exit 0;
\end{lstlisting}
    \caption{A script with an invalid expression (line~\ref{l2:i:error}) produced by generator~\genI.}
    \label{fig:example2:I}
  \end{subfigure}
  \begin{subfigure}[b]{0.365\linewidth}
\begin{lstlisting}[numbers=left]
for E9 in D a; (*\label{l2:mi:error}*)
 Ldn="$((3&4))" && echo "$Ldn";
done
exit 0;
\end{lstlisting}
    \caption{A script with a syntax error (line~\ref{l2:mi:error}) produced by generator \genM.}
    \label{fig:example2:MI}
  \end{subfigure}
\caption{Three shell scripts generated by \shf. }
\label{fig:example2-scripts}
\end{figure}

\autoref{fig:example2-scripts} shows examples of scripts produced by the three generators.  Generator \genV produces the script in \autoref{fig:example2:V}, which is in fact syntactically valid according to the POSIX standard; therefore, executing the script should trigger no interpreter errors, and it should terminate by returning \Sh{0} as exit code.  Generator \genI produces the script in \autoref{fig:example2:I}, which includes the invalid test command \Sh{[ - done func87();} -- which is invalid because there is no closing square bracket and keyword \Sh{done} cannot be used in a test construct; therefore, executing the script should result in an error when evaluating the expression.  Generator \genM generates the script in \autoref{fig:example2:MI}, which is an invalid variant of the valid \autoref{fig:example2:V} -- invalid because it elided the required \Sh{do} keyword; therefore, executing the script should also result in an error when evaluating the loop.  In this case, the invalid expression was generated by applying a mutation (removing \Sh{do} keyword) to the initially valid script of \autoref{fig:example2:V}.

\begin{table}[!bt]
  \centering
  \begin{tabular}{llc}
    \toprule
    \multicolumn{1}{c}{\textsc{bug category}} 
    & \multicolumn{1}{c}{\textsc{failure}}
    & \multicolumn{1}{c}{\textsc{generators}}
    \\
    \midrule
    \multirow{2}{*}{logic error}
    & $\mathtt{ec} \neq 0$, error message
    & \genV
    \\
    & $\mathtt{ec} = 0$, no error message
    & \genI, \genM
    \\ \cmidrule(lr){2-3}
    \multirow{2}{*}{memory error}
    & address sanitizer error message
    & \genV, \genI
    \\
    & memory sanitizer error message
    & \genV, \genI
    \\ 

		\bottomrule
\end{tabular}
   
\caption{Categories of bugs detected by \shf. For each \textsc{bug category}, the \textsc{failure} that signals the bug, and which \textsc{generators} may find such bugs.}
  \label{tab:bug_cat}
\end{table}

\subsubsection{Bug Categories}

The shell scripts generated by \shf can reveal various categories of bugs in shell interpreters, as summarized in \autoref{tab:bug_cat}.  As explained, each generator produces valid (\genV) or invalid (\genI, \genM) scripts with an automatically generated oracle that expects the absence (\genV) or presence (\genI, \genM) of errors (non-zero exit code, or specific messages printed on standard output/error).  Whenever there is a failure of one of such oracles, it may reveal a \emph{logic} error -- that is, an incorrect behavior of the shell interpreter.  For example, if a POSIX-compliant script generated by \genV triggers a syntax error, it may indicate that the shell's parser has a bug.

\sloppy{In order to broaden the bug-detection capabilities of \shf, we equipped its test-execution environment with address~\cite{addresssanitizer} and memory leak~\cite{memorysanitizer} sanitizers: whenever any of the scripts generated by \shf executes, the sanitizers monitor the execution environment for various kinds of memory errors.}  Therefore, \shf treats the sanitizers' error messages as additional failures, which may reveal various kinds of \emph{memory} errors, including \emph{out-of-bound} memory accesses (e.g., a buffer overflow), \emph{pointer} errors (e.g., a null-pointer dereference), and \emph{memory leaks} (e.g., when heap-allocated memory is not released after use).

\subsection{Implementation Details}
\label{sec:implementation}

We implemented the \shf technique in a prototype tool with the same name.  At its core, \shf uses \AFL's grammar mutator~\cite{gram_mutator,nautilus2} with the three grammars for \genV, \genI, \genM that we designed as outlined in \autoref{sec:methodology}.  In contrast to how it is used within a fuzzer like \AFL, \shf uses productions of the grammars that are free from random mutations.  The only exception is generator \genM, which may add single-location mutations to its generated scripts; however, we carefully designed \genM's mutation operators so that they only hamper the (syntactic) correctness of the generated scripts, and are unlikely to introduce destructive behavior.

\shf is written in a combination of C, C++, Python and shell scripts. It includes functions to prepare the environment and compile the grammars, an interface to \AFL's grammar mutator, an execution engine that orchestrates the testing process, as well as logging and post-processing utilities that are written in Python and shell scripts.

\shf's current implementation does not run sanitizers with generator \genM (whereas it runs them with \genV and \genI).  This is just a simplification of the implementation: \genV and \genI already cover a broad spectrum of behaviors that may trigger memory errors, whereas \genM mainly brings additional diversity to detect different kinds of \emph{logic} bugs.

\begin{table}[!bt]
  \centering
  \begin{tabular}{llc}
    \toprule
    \multicolumn{1}{c}{\textsc{parameter}}
    & \multicolumn{1}{c}{\textsc{description}}
    & \multicolumn{1}{c}{\textsc{generators}}
    \\
    \midrule
    $\param{size}$
    & approximate size (LOC) of generated scripts
    & \genV, \genM
    \\
    $\param{procs}$
    & number of parallel shell processes
    & \genV, \genI, \genM
    \\
    $\param{time}$
    & time allotted to the generator
    & \genV, \genI, \genM
    \\
	\bottomrule
  \end{tabular}
   
\caption{Main parameters used to configure \shf's executions}
  \label{tab:params_desc}
\end{table}

\subsubsection{Parameters}
\label{sec:parameters}

Users can customize \shf's runs by setting the three main parameters listed in \autoref{tab:params_desc}.

\begin{description}
\item[\param{size}:] the approximate size, in lines of code, of each generated script. The size is ``approximate'' because \AFL's grammar mutator, on top of which \shf is built, does not enforce strict constraints on the generated program size. Also, parameter \param{size} is not applicable to generator \genI, since the sizes of the produced test cases are already constrained by the grammar itself. 

\item[\param{procs}:] the number of parallel shell processes that will run the generated scripts.  In order to improve throughput, \shf runs parallel instances of the generators, whose produced scripts are fed into parallel instances of the shell interpreter under test.

\item[\param{time}:] the overall time available for generating shell scripts, executing them, and collecting the results.
\end{description}

Users can set these parameters according to the goals of their testing campaign.  In \autoref{sec:results:rq4}, we will discuss how we empirically selected \shf's parameters to use in the bulk of our experimental evaluation of its effectiveness.

\begin{figure}[!tb]
  \centering
  \small
\begin{algorithmic}
{\normalsize \PROCEDURE{\proc{reduce}}{\param{script}, \param{oracle}, \param{n}}{\param{reduced script}}}
\TRUE

\IF {\proc{size}(\param{script}) = 1}
    \STATE \COMMENT {\param{script} cannot be split into smaller chunks}
    \RETURN \param{script}
\ELSE
    \STATE \COMMENT {split \param{script} into \param{n} chunks}
    \STATE \param{chunks} $\leftarrow$ \proc{split}(\param{script},\param{n})
    \FOR {\param{chunk} $\in$ \param{chunks}}
       \STATE \COMMENT {remove \param{chunk} from \param{script}}
       \STATE \param{shrunk} $\leftarrow$ \param{script} $-$ \param{chunk}
       \IF {\param{oracle}(\param{shrunk}) = \param{oracle}(\param{input})}
            \STATE \COMMENT {\param{shrunk} preserves \param{input}'s behavior according to \param{oracle}}
            \STATE \COMMENT {try to further reduce it}
            \RETURN \proc{reduce}(\param{shrunk}, \param{oracle}, max($\param{n}-1$, $2$))
        \ENDIF
    \ENDFOR
    \STATE \COMMENT {none of the shrunk scripts preserves \param{input}'s behavior}
    \IF {\param{n} < \proc{size}(\param{script})}
        \STATE \COMMENT {try using smaller chunks}
        \RETURN \proc{reduce}(\param{script}, \param{oracle}, min($2\times\param{n}$, \proc{size}(\param{script})))
     \ELSE
        \STATE \COMMENT {smallest chunk size reached: stop}
        \RETURN \param{script}
    \ENDIF
\ENDIF   
\end{algorithmic}

\caption{Delta-debugging algorithm used to reduce scripts generated by \shf.}
\label{fig:reduct_algorithm}
\end{figure}

\subsubsection{Script Size Reduction}
\label{sec:minimization}

As we'll demonstrate in \autoref{sec:exp-results}, \shf may generate scripts of non-trivial size -- up to hundreds of lines of code.  While a variety in the sizes of generated scripts improves metrics such as coverage, whenever a script triggers a failure it has to be inspected to determine whether the failure actually corresponds to a genuine bug.  In such cases, we would ideally only have to inspect scripts in \emph{reduced} form: as small as possible, while still triggering the same failure.  To this end, we added to \shf an optional script reduction\footnote{While it is customary to use the term ``minimization'', our algorithm is not guaranteed to find the smallest possible equivalent test case -- even though it is quite effective in practice.} step.

As customary in test-case generation~\cite{testcase_reduct_survey}, we used the delta debugging~\cite{delta_debugging} algorithm shown in \autoref{fig:reduct_algorithm}.  Our implementation uses a function \proc{split} that is aware of the syntactic structure of shell scripts.  Furthermore, for invalid scripts generated by \genI and \genM, the \proc{split} function also avoids removing parts of any syntactically invalid element; and, for valid scripts generated by \genV, it avoids splitting the statement that triggers the (unexpected) failure.  Consequently, a script produced by a certain generator preserves its characteristics (valid or invalid by construction) even after reduction.  \autoref{fig:example_reduction_v} shows an example of reducing a script generated by \genV.\footnote{This is one of the memory leaks detected by \genV.} The script was originally over 500 lines of code, and triggered a failure at line~\ref{l:failure-reduction}; the reduction algorithm shrunk it down to a single failing assignment.  As you can see in \autoref{fig:example_reduction_v}, \proc{split} looks for chunks that consist of complete statements, and does not further break down the failure-inducing statement at line~\ref{l:failure-reduction}.

\begin{figure}[!tb]
  \centering
  \begin{tikzpicture}
    \node (step1) {
\begin{lstlisting}[numbers=left,firstnumber=1]
(*\highlighcode{if [ ! Elb '>' "qhx" ]; then}*)
(*\highlighcode{    B9=3}*)
(*\highlighcode{fi}*)

# failing line:
b="$((E/t))" du && b0="${BnE}"(*\label{l:failure-reduction}*)
\end{lstlisting}};
    \node[right=of step1] (step2) {
\begin{lstlisting}[numbers=left,firstnumber=1]




# failing line:
b="$((E/t))" du (*\highlighcode{\&\& b0="\$\{BnE\}"}*)
\end{lstlisting}};
    \node[right=25mm of step2] (step3) {
\begin{lstlisting}[numbers=left,firstnumber=1]




# failing line:
b="$((E/t))" du 
\end{lstlisting}};

    \begin{scope}[->,shorten >=4mm,very thick,darkbluecol]
      \draw ($(step1.east)+(-5mm,0)$) -- (step2);
      \draw ($(step2.east)+(10mm,0)$) -- (step3);
    \end{scope}
  \end{tikzpicture}
    \caption{Three final steps of the \autoref{fig:reduct_algorithm}'s script reduction procedure, demonstrated on a program generated by \genV. The chunk of code that is removed in each reduction step is highlighted.}
  \label{fig:example_reduction_v}

\end{figure}

\section{Experimental Design}
\label{sec:exp-design}

We designed an experimental evaluation of \shf to address the following research questions:

\begin{description}

\item[RQ1:] 
  Is \shf effective in detecting bugs in shell interpreters?
  \\
  This research question investigates \shf's fundamental effectiveness as a bug-finding tool for shell interpreters.
  
\item[RQ2:] 
  Which categories of bugs can \shf detect?
  \\
  This research question looks into the categories of shell-interpreter bugs detected by \shf, and into which generators are more/less effective.
  
\item[RQ3:] How does \shf compare with the state-of-the-art fuzzing testing framework \AFL?
  \\
  \sloppy{This research question compares \shf to using an off-the-shelf fuzzer like \AFL (also equipped with a grammar to guide the generation of meaningful shell scripts).}  Among the many fuzzing frameworks that are available, we consider \AFL~\cite{aflpp} because it is one of the most actively maintained, recent forks of the influential AFL,\footnote{\url{https://lcamtuf.coredump.cx/afl/}} which implements several improvements, variants, and extensions that have been proposed in the literature.
  
\item[RQ4:] What is an effective set of parameters to configure \shf?
  \\
  This research question investigates the tuning of \shf's configuration parameters (described in~\autoref{sec:implementation}), and determines a suitable configuration that was used in all other research questions.

\end{description}

The rest of this section describes the experiments we designed to answer each research question.

\subsection{Shell Interpreters}
Each experiment ran in a virtual machine with one core and 15~GB of RAM and Ubuntu 20.04.
 
As shell interpreters, we primarily targeted version~R59 of the POSIX-com\-pli\-ant shell \mksh.\footnote{MirBSD: mksh -- the MirBSD Korn Shell: \url{http://mirbsd.de/mksh}} \mksh is a modern variant of KornShell (\verb+ksh+), originally developed for the MirOS BSD system.  Compared to other, longer-running POSIX-com\-pli\-ant shell interpreters such as Bash, the C shell, and the original Korn Shell, \mksh is less widely used and is actively maintained by a single programmer;\footnote{\url{https://launchpad.net/~mirabilos}} thus, it is a promising target for discovering new, previously unknown bugs and issues through fuzzing.  \mksh is also the default shell on Android, and as such it is an important component of countless mobile devices; thus, its correctness and security also indirectly affect the robustness of these systems -- which further motivates our effort to automatically identify bugs and vulnerabilities.

Even though our experiments focused on \mksh, the shell scripts produced by \shf work equally well as tests of other shell interpreters.  Whenever \shf generated a script that resulted in an alert when executed by \mksh, we ran the same script also with other common shell interpreters (Bash, KornShell, and Dash) to ascertain whether they are also affected by the same issue.  \autoref{sec:exp-results} will also report any such cases of warning replicated with other shells.

\subsection{Parameters}
\autoref{tab:parameters-default}
shows the default parameters we used in all experiments (except those for RQ4).  We selected these parameters empirically, as a result of the experiments we ran to answer RQ4 (see \autoref{sec:results:rq4}).

\begin{table}[!bt]
  \centering
  \begin{tabular}{c rrr}
    \toprule
    & \multicolumn{3}{c}{\textsc{parameters}}
    \\ \cmidrule(lr){2-4}
    \multicolumn{1}{c}{\textsc{generator}}
    & \multicolumn{1}{c}{\param{size}}
    & \multicolumn{1}{c}{\param{procs}}
    & \multicolumn{1}{c}{\param{time} [h]}
    \\
    \midrule
    \genV & 500 & 100 & 4 
    \\
    \genI &  -- & 1000 & 6
    \\
    \genM & 50 & 1000 & 2
    \\
    \bottomrule
  \end{tabular}
  \caption{Parameter values used in the main experiments with \shf. Parameter \param{time} is measured in hours, and refers to when the three generators are used sequentially within the same 12-hour run.}
  \label{tab:parameters-default}
\end{table}

Each experiment ran \shf or, in RQ3, \AFL for a total of 12 hours.  To account for random fluctuations, we repeated each experiment three times, and report average values.

Since \shf's generators build scripts with complementary characteristics, in most experiments we ran all three generators sequentially \genV, \genI, and \genM within the same 12-hour budget, and collected all results.  \autoref{tab:parameters-default} shows the amount of time given to each generator within a 12-hour run (which we also determined empirically when answering RQ4).

\subsection{Measures}
In each experiment, we executed all scripts generated by \shf with the \mksh shell interpreter instrumented with address and memory sanitizers.  For each script execution, we logged:
\begin{description}
\item[generator:] the generator that produced the script
\item[alert:] whether the execution raised an \emph{alert}:
  either a failure of the generator's built-in oracle,
  or a sanitizer warning (see \autoref{tab:bug_cat})
\item[size:] the script's size, in lines of code
\end{description}
If a script execution raises an alert, we also \emph{save} the script itself for post processing.  Furthermore, we also record the maximum \emph{branch} \textbf{coverage} reached during the whole experiment.

\subsection{Post processing}
\label{sec:post-processing}
After an experiment concludes, we automatically group all saved scripts that triggered \emph{alerts}:
any two alerts belong to the same group if and only if they were produced by the same generator, they trigger the same oracle failure or sanitizer warning, and determine the same shell exit code and error message.\footnote{ For invalid scripts produced by generators \genI or \genM, the ``error message'' may be a sanitizer warning, or denote the absence of an expected shell interpreter error.  }

After this grouping of alerts into \emph{unique} alerts, we take the smallest script in each group and \emph{reduce} its size as described in \autoref{sec:minimization}.  Finally, we manually inspect each unique, reduced alert, in order to determine whether it corresponds to a genuine \emph{bug} (i.e., a true positive).  During the manual inspection, we first of all determine whether the script is indeed POSIX-compliant (if produced by generator \genV) or invalid (if produced by generators \genI or \genM); then, we ascertain that it does trigger unexpected behavior according to the standard.

For all unique alerts that we classified as true positives, we also prepared a bug report and submitted it to \mksh's maintainers.  Following standard guidelines \cite{bug_report_guidelines}, a bug report includes:
\begin{enumerate*}
\item the source code of the bug-revealing shell script;
\item information about the hardware, operating system environment, and shell interpreter used in our experiments that revealed the bug;
\item a short description of the expected script behavior, in contrast to the (unexpected) behavior we observed in the experiments;
\item detailed steps to reproduce the observed behavior
  (including the stack trace in case of a crashing bug);
\item if possible, a brief discussion of how to address the issue,
  such as a suggested code or documentation change.
\end{enumerate*}
To avoid publicly disclosing potential security issues, we submitted all bug reports as encrypted emails sent directly to the maintainers of the shell interpreter.  If the maintainers acknowledged the bug report as a genuine issue that should be fixed, we classify the corresponding bug as \emph{confirmed}.  Confirmed bugs are the gold standard for the effectiveness of a test generation technique, as they represent authentic software defects.

\section{Experimental Results}
\label{sec:exp-results}

This section presents the experimental results that pertain to each research question.

  \begin{table}[!bt]
    \setlength{\tabcolsep}{3pt}
    \small
    \centering
  \begin{tabular}{c l rr rr rr rr}
    \toprule
    & & \multicolumn{8}{c}{\textsc{generators}}
    \\
    \cmidrule(lr){3-10}
    & 
    & \multicolumn{2}{c}{$\genV + \genI + \genM$}
    & \multicolumn{2}{c}{\genV}
    & \multicolumn{2}{c}{\genM}
    & \multicolumn{2}{c}{\genI}
    \\
    \multicolumn{2}{c}{\textsc{category}}
    & \multicolumn{1}{c}{\textsc{sum}} & \multicolumn{1}{c}{\textsc{avg}}
    & \multicolumn{1}{c}{\textsc{sum}} & \multicolumn{1}{c}{\textsc{avg}}
    & \multicolumn{1}{c}{\textsc{sum}} & \multicolumn{1}{c}{\textsc{avg}}
    & \multicolumn{1}{c}{\textsc{sum}} & \multicolumn{1}{c}{\textsc{avg}}
    \\
    \cmidrule(){1-2}
    \cmidrule(l){3-4}
    \cmidrule(l){5-6}
    \cmidrule(l){7-8}
    \cmidrule(l){9-10}
    
    \multirow{4}{*}{\textsl{all}}
    &
        \textsl{alerts} & 203460 & 67820.0 
                     & 930 & 310.0 
                     & 4339 & 1446.3 
                     & 401583 & 133861.0 
        \\
        & \textsl{unique}  & 11 & 10.3 
                & 2 & 2.0 
                & 4 & 3.7 
                & 8 & 6.7 
        \\
        & \textsl{true positives} & 8 & 8.0 
                      & 2 & 2.0 
                      & 1 & 1.0 
                           & 7 & 5.7 
        \\
        & \textsl{confirmed}  & 7 & 7.0 
                   & 2 & 2.0 
                   & 1 & 1.0 
                           & 6 & 4.7 
    \\
    \cmidrule(){1-2}
    \cmidrule(l){3-4}
    \cmidrule(l){5-6}
    \cmidrule(l){7-8}
    \cmidrule(l){9-10}
    \multirow{4}{*}{\textsl{logic}}
    &
        \textsl{alerts} & 203104 & 67701.3 
                     & 0 & 0.0 
                     & 4339 & 1446.3 
                     & 401497 & 133832.3 
                     
        \\
                       
        & \textsl{unique} & 9 & 8.3 
                & 0 & 0.0 
                & 4 & 3.7 
                & 6 & 6.0 
        \\
        & \textsl{true positive}  & 6 & 6.0 
                       & 0 & 0.0 
                       & 1 & 1.0 
                           & 5 & 5.0 
        \\
        & \textsl{confirmed}  & 5 & 5.0 
                   & 0 & 0.0 
                   & 1 & 1.0 
                   & 4 & 4.0 
    \\
    \cmidrule(){1-2}
    \cmidrule(l){3-4}
    \cmidrule(l){5-6}
    \cmidrule(l){7-8}
    \cmidrule(l){9-10}
    
    \multirow{4}{*}{\textsl{memory}}
    &
        \textsl{alerts}  & 356 & 118.7 
                    & 930 & 310.0 
                    & 0 & 0.0 
                    & 86 & 28.7 
        \\
        & \textsl{unique} & 2 & 2.0  
               & 2 & 2.0  
               & 0 & 0.0  
               & 2 & 0.7  
        \\
        & \textsl{true positive} & 2 & 2.0 
                      & 2 & 2.0 
                      & 0 & 0.0 
                      & 2 & 0.7 
        \\
        & \textsl{confirmed} & 2 & 2.0 
                  & 2 & 2.0 
                  & 0 & 0.0 
                  & 2 & 0.7 
    \\
    \cmidrule(){1-2}
    \cmidrule(l){3-4}
    \cmidrule(l){5-6}
    \cmidrule(l){7-8}
    \cmidrule(l){9-10}
    
    \multicolumn{2}{c}{\textsc{coverage}}
    &  \multicolumn{2}{c}{37.9 \%} 
                 &  \multicolumn{2}{c}{29.7 \%} 
                 &  \multicolumn{2}{c}{27.6 \%} 
                 &  \multicolumn{2}{c}{37.3 \%} \\ 
    \bottomrule
	\end{tabular}
   \caption{Effectiveness of \shf.
     The table reports the total number \textsc{sum}
     over three repeated 12-hour runs, and average \textsc{avg} per repeated run
     of \textsl{alerts} raised by \shf,
     \textsl{unique} alerts,
     \textsl{true positive} bugs,
     and bugs \textsl{confirmed} by the shell maintainers.  The three row groups report data for \textsl{all} alerts, as well as the breakdown into \textsl{logic} and \textsl{memory} alerts (identified as described in \autoref{sec:methodology}).  The four column groups report data for all three generators used sequentially
     ($\genV + \genI + \genM$),
     as well as for each generator used individually in a run.  Finally, the bottom row reports the maximum branch \textsc{coverage} reached during any run.  }
 \label{tab:summary_generators_v2}
 \end{table}

 \begin{figure}[!bt]
   \centering
   \begin{tikzpicture}[ultra thick]
     \coordinate (L) at (0,0);
     \coordinate (R) at (10mm,0);
     \coordinate (B) at ($(L)!0.5!(R)+(0,-10mm)$);
     \draw[aflcol] let \p1 = ($(R)-(L)$), \n2 = {veclen(\x1,\y1)}
     in (L) circle (\n2) node [label={[align=center,label distance=8mm]170:\genV\\[-2pt]2}] {};
     \draw[kleecol] let \p1 = ($(R)-(L)$), \n2 = {veclen(\x1,\y1)}
     in (R) circle (\n2) node [label={[align=center,label distance=8mm]10:\genI\\[-2pt]7}] {};
     \draw[cbmccol] let \p1 = ($(R)-(L)$), \n2 = {veclen(\x1,\y1)}
     in (B) circle (\n2) node [label={[align=center,label distance=8mm]-30:\genM\\[-2pt]1}] {};
     \node at ($(L)+(-12pt,7pt)$) {0};
     \node at ($(R)+(12pt,7pt)$) {5};
     \node at ($(B)+(0pt,-10pt)$) {1};
     \node at ($(L)!0.5!(R)+(0,10pt)$) {2};
     \node (X) at ($(L)!0.5!(R)+(0,-10pt)$) {0};
     \node at ($(X)+(-18pt,-8pt)$) {0};
     \node at ($(X)+(18pt,-8pt)$) {0};
   \end{tikzpicture}
   \caption{Out of all 8 unique true positive alerts observed in \shf's experiments, which were raised by scripts generated by each of the generators \genV, \genI, and \genM.}
   \label{fig:uniqueness}
 \end{figure}

\subsection{RQ1: Effectiveness}
\label{sec:results:rq1}

In our experiments, the shell scripts generated by \shf raised a large number of alerts; several of them correspond to true positives that indicate bugs or other kinds of issues in the shell interpreters.  As shown in \autoref{tab:summary_generators_v2}, in the experiments where \shf ran all three generators sequentially
($\genV + \genI + \genM$),
\shf triggered over \numprint{200000}
alerts; after post processing (\autoref{sec:post-processing}), these reduce to 11 unique \emph{alerts}, 8 of which correspond to \emph{true positives} -- all but one \emph{confirmed} by the maintainers of \mksh as genuine issues of the shell interpreters that should be fixed.\footnote{ We thank Thorsten Glaser, the maintainer of \mksh, for his kind availability and promptness in answering and addressing our issue reports.  }

It is typical for such experiments that only a small number of unique alerts emerge from a much larger number of (raw) alerts.  Thus, \shf's automatic grouping of alerts into unique alerts (which, in turn, is based on the \emph{oracles} that each generated test script is equipped with) is crucial to reduce the amount of output that the user needs to look at.  Generating a much larger number of scripts still remains useful to increase the \emph{coverage} of the testing process: a coverage of over 30\% indicates that \shf exercised a considerable portion of the shell interpreter under test, which bolsters the significance of the issues it exposed.

In \autoref{tab:summary_generators_v2}, the total number of unique, true positive, and confirmed alerts is usually close -- often identical -- to the average number per run.  This means that the same issues are often revealed in most -- or all -- of the repeated runs; it indicates that \shf is \emph{robust} as a test generator: even though its generation process is randomized, it often manages to perform a broad, repeatable search if it is allowed to run for several hours.

\begin{figure}[!bth]
  \centering
  \begin{subfigure}[b]{0.48\linewidth}
\begin{lstlisting}[numbers=none,escapechar={|}]
        read VAR && echo fi;
\end{lstlisting}
    \caption{A script raising a spurious alert.}
    \label{fig:example:unique-fp}
  \end{subfigure}
  \begin{subfigure}[b]{0.51\linewidth}
\begin{lstlisting}[numbers=none]
shift -- -0000000000000000000000000000000000;
\end{lstlisting}
    \caption{A script raising a true positive alert, not confirmed by developers.}
    \label{fig:example:tp-notconfirmed}
  \end{subfigure}
\caption{Two shell scripts generated by \shf that raised unique alerts.}
\label{fig:example-rq1}
\end{figure}

\subsubsection{Precision}

\shf's overall precision is 72\%, that is the fraction $8/11$ of unique alerts that were confirmed as true positives.  Given that the absolute number of unique alerts is modest, this precision level is adequate, as it does not require the user to inspect and identify a large number of false positives.

In the few cases where \shf generated false positives, they were usually due to unusual combinations of syntactically invalid constructs that unexpectedly resulted in valid programs.  For example, \autoref{fig:example:unique-fp} shows a script that raised a unique alert that is false positive.  The script was generated by \genM, which removed parts of a conditional block, leaving only the \Sh{fi} closing keyword.  The mutation removing parts of a conditional is supposed to render the script \emph{invalid}; however, \autoref{fig:example:unique-fp}'s script is perfectly valid, since \verb+fi+ is simply interpreted as a string printed by \Sh{echo}.  Since the script was generated by \genM, its oracle expects an error; the script executes correctly, and hence it results in an alert; but since the script is actually valid, the alert is spurious (false positive).

\autoref{fig:example:tp-notconfirmed}
shows the only case of script that raised a true positive alert, which was not confirmed by the maintainers of \mksh.  The script was generated by \genI: according to the POSIX standard, the argument of \Sh{shift} ``shall be an unsigned decimal integer'',\footnote{\smaller\url{https://pubs.opengroup.org/onlinepubs/9699919799/utilities/V3_chap02.html\#shift}} and hence the signed \Sh{-0} should be rejected.  However, shells usually accept a signed zero as equivalent to an unsigned one; according to the maintainers of \mksh, this behavior has been standard in Unix shells; thus, they decided against disallowing this behavior even if it is not strictly POSIX-compliant.

These examples demonstrate a tension between standard-compliance and legacy behavior that is common in all widely-used real-world languages: if a certain behavior has been historically allowed, maintainers are reluctant to disallow it (even when it violates a standard) because doing so would introduce -- with no apparent advantage -- unexpected errors (or undefined behavior) in programs that users would consider valid.

\subsubsection{Generators}

If we break down \shf's results by generator (middle and right columns in \autoref{tab:summary_generators_v2}) we notice that the three generators are mostly \emph{complementary}, in that each of them leads to the discovery of unique alerts that the other generators did not detect.  However, the generators are not perfectly complementary: as shown in \autoref{fig:uniqueness}, some unique alerts can be triggered by the scripts produced by different generators.  This is not surprising given that:
\begin{enumerate*}
\item all experiments use sanitizers, and hence every generators may be able to trigger (the same) memory errors (this is the case, in particular, of the two memory errors detected by both \genI and \genV);
\item generators \genI and \genM both generate \emph{invalid} shell scripts,
  which may have the same source of invalidity.
\end{enumerate*}
Even though it may seem that generator \genV's effectiveness is subsumed by generator \genI's, generator \genV's detection of memory errors behaved more robustly in our experiments: only \genV detected the two errors in all repeated runs
(in \autoref{tab:summary_generators_v2}, $\textsc{sum} = \textsc{avg}$ for \genV but not for \genI).

The generators also differ in the coverage they achieve, with \genI reaching the highest coverage and \genM the lowest.  Indeed, \genM produces scripts with a fairly predictable, constrained structure, that mainly differ in size, and in the mutations they include; this restricts the coverage it obtains.  In contrast, \genI produces scripts with a variety of invalid syntactic features, which leads to exercising a range of error-checking code in the shell interpreters.

Remember that \autoref{tab:summary_generators_v2}'s results under columns \genV, \genI, and \genM are not merely a breakdown of those under column $\genV + \genI + \genM$, but correspond to experiments where we ran each generator alone for the whole allotted time.  It is interesting that no generator discovered issues when run in isolation that it did not also discover when sharing the testing time with the other generators.  This is additional evidence that \shf's runs are generally robust, and its three generators are largely complementary and thus are best run together.

\begin{table}[!hb]
  \centering
  \begin{tabular}{lcrrr}
    \toprule
    & & \multicolumn{3}{c}{\textsc{generator}}
    \\
    \multicolumn{1}{c}{\textsc{scripts}}
    & \multicolumn{1}{c}{\textsc{reduced?}}
      & \genV & \genM & \genI 
    \\
    \midrule
    \textsl{all} & \suppNO & 694 & 56 & 5 \\[2mm]
    \textsl{alerts} & \suppNO  & 768 & 2 & 2 \\[2mm]
    \multirow{2}{*}{\textsl{true positives}} & \suppNO  & 566 & 1 & 4 \\
    & \suppOK & 4 & 1 & 1  \\
    \bottomrule
  \end{tabular}
  \caption{Average size, in lines of code, of the scripts generated by \shf
    in RQ1's experiments: \textsl{all} scripts, those raising \textsl{alerts}, and the alerts confirmed as \textsl{true positives} (both before \suppNO\ and after \suppOK\ size reduction).}
  \label{tab:alerts_sizes}
\end{table}

\begin{figure}
\begin{subfigure}{0.49\textwidth}
\centering
\includegraphics[width=1.0\linewidth]{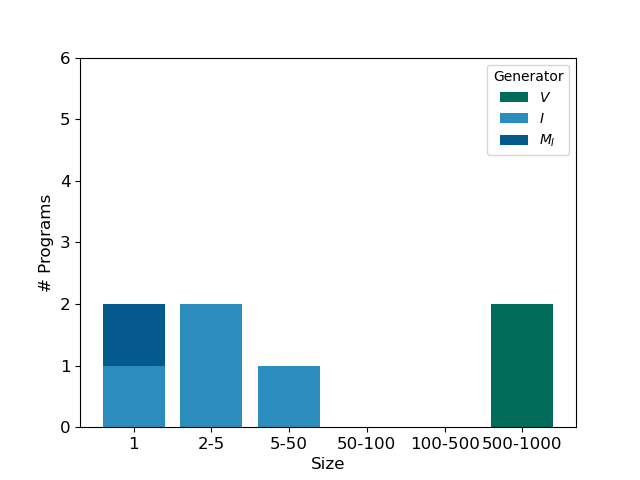}
\caption{Distribution of sizes before reduction.}
\end{subfigure}
\begin{subfigure} {0.49\textwidth}
\includegraphics[width=1.0\linewidth]{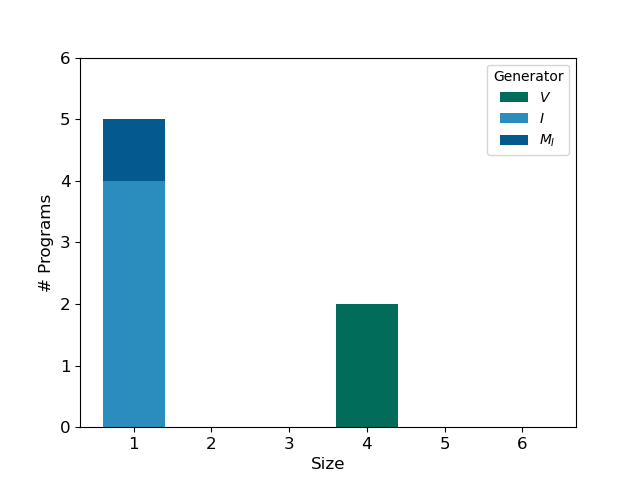}
\caption{Distribution of sizes after reduction.}
\end{subfigure}
\caption{Distribution of the size of all script generated in RQ1's experiments that raised true positive alerts, before and after the reduction step described in \autoref{sec:minimization}.}
\label{fig:histogram_putsize}

\end{figure}

\subsubsection{Size}

As shown in \autoref{tab:alerts_sizes}, the three generators differ significantly in the \emph{size} of scripts they generate.  Generator \genV produces by far the largest scripts, usually around 600--700 lines of code; generator \genI produces the smallest scripts, usually around 5 lines of code; generator \genM is in between, as it usually produces scripts of around 50--60 lines of code.

\autoref{tab:alerts_sizes} and \autoref{fig:histogram_putsize} also show that \shf's post processing (\autoref{sec:post-processing}) is crucial to produce scripts of smaller size, which can easily be manually inspected.  The script reduction algorithm described in \autoref{sec:minimization} is essential for scripts generated by \genV, which it shrinks by two order of magnitudes.  In contrast, it is less essential for scripts generated by \genM and \genI, where the smallest script in each equivalence group is usually only a few lines long; however, the reduction algorithm manages to shrink these even further, bringing them down to a single line of code that clearly highlights the root cause of the bug in each case.

\finding{
  In our experiments, the scripts produced by \shf detected 7 confirmed issues in the \mksh shell interpreter, while achieved a coverage of over 30\%.  \shf's generation is often robust (the same issues are detected in different repeated runs).  \shf's three generators \genV, \genI, and \genM are largely complementary, as they revealed different issues.  }

\begin{figure}[!bt]
  \centering
  \begin{subfigure}[b]{0.32\linewidth}
\begin{lstlisting}[numbers=left]
b=0;
if [ ! -f file1 ]; then
  echo "$USER" > file1;
  b="$((E/t))" du
fi
exit 0;
\end{lstlisting}
    \caption{A POSIX-valid script, produced by generator \genV.}
    \label{fig:example:V}
  \end{subfigure}
  \begin{subfigure}[b]{0.32\linewidth}
\begin{lstlisting}[numbers=left]
i=3;
if ((3+"i")); then
  a="$((3+"i"))"(*\label{l:i:error}*)
  echo "${a}"
fi
\end{lstlisting}
    \caption{A script with an invalid expression (line~\ref{l:i:error}) produced by generator \genI.}
    \label{fig:example:I}
  \end{subfigure}
  \begin{subfigure}[b]{0.32\linewidth}
\begin{lstlisting}[numbers=left]
while [ "$st" = "" ]; do
  if [ true ]; then
    if [ true ]; then
      mksh -c "((3+3"(*\label{l:mi:error}*)
    fi
  fi
done
\end{lstlisting}
    \caption{A script with a syntax error (line~\ref{l:mi:error}) produced by generator \genM.}
    \label{fig:example:MI}
  \end{subfigure}
\caption{Three shell scripts generated by \shf that trigger confirmed bugs.}
\label{fig:example-scripts}
\end{figure}

\subsection{RQ2: Categories of Bugs}
\label{sec:results:rq2}

As we discussed in \autoref{sec:methodology}, \shf's generators are specifically designed to detect different kinds of \emph{logic} errors in shell interpreters.  Indeed, \autoref{tab:summary_generators_v2} shows that the vast majority of alerts, and 9 out of the 11 unique alerts, raised in our experiments correspond to logic errors.  Despite its focus on logic errors, \shf also detected 2 unique memory errors, both confirmed by developers.  This suggests that running sanitizers while executing \shf's scripts is worth the (minor) performance overhead, since it broadens the number and kinds of errors that \shf can detect.

In terms of generators, all logic errors were detected by programs produced by generators \genI and \genM.  For instance, \genI produced the script in \autoref{fig:example:I}, which exposes one of the 5 confirmed unique logic errors we found in our experiments.  \autoref{fig:example:I}'s script is invalid because it includes twice the expression \Sh{((3+"i"))} (invalid because it combines a number and a string in an arithmetic expression); \mksh, however, only gives a warning that does not prevent execution from continuing after evaluating the invalid expression.  \autoref{fig:example:MI} shows another example of a script -- produced by \genM -- that exposed a confirmed unique logic error.  Also in this case, \mksh executes the program without errors or warnings, even though expression \Sh{((3+3} is clearly invalid because its parentheses are not properly balanced.

Interestingly, generator \genV only led to discovering memory errors.  In other words, none of the valid scripts it generated were rejected by the shell interpreters.  For instance, \genV produced the script in \autoref{fig:example:V}, which executes without explicit error but triggers one of the 2 confirmed unique memory errors we found in our experiments.  The memory errors exposed by generator \genV were also the only errors in our experiments that also affected version~5.0 of Bash -- they have been fixed since version~5.2.\footnote{ In contrast, all logic errors found in our experiments only affected \mksh.  } In all, generator \genV remains useful to exercise the shell interpreters in different ways, as well as to improve coverage; and further demonstrates that \shf's three generators have \emph{complementarity} capabilities.

\finding{
  In our experiments, \shf was especially effective in detecting 5 confirmed logic errors, but could also identify 2 confirmed memory errors that affected both \mksh and \texttt{bash}.  Using all generators \genV, \genI, and \genM increases the variety of bugs that \shf reveals.
}

\subsection{RQ3: Comparison}
\label{sec:results:rq3}

We compare \shf to the state-of-the-art fuzzer \AFL\footnote{Our comparison targeted \AFL version~4.05.} on testing the \mksh shell interpreter.

\subsubsection{Setup}
To make the comparison fair, we equip \AFL with its grammar mutator~\cite{gram_mutator} and the same grammar of valid POSIX shell scripts used by generator \genV.  Even though \shf's implementation also uses \AFL's grammar mutator in its implementation, the scripts produced by \AFL and \shf's \genV will be in general quite different.  In \AFL, the fuzzing engine may always introduce some random, unconstrained mutations on top of the seed scripts produced according to \genV's grammar; therefore, there is no guarantee that any of the scripts produced by \AFL with the grammar mutator are valid.

Consequently, we cannot associate a ``validity'' oracle to the shell scripts produced by \AFL, since we do not know which are valid and which invalid according to the POSIX standard.  As it is customary in fuzzing campaigns, we simply use memory sanitizers to detect memory-related errors that are revealed when running such scripts.

\AFL is a highly configurable tool that supports different search strategies; we consider three different configurations $A_1, A_2, A_3$ for it.  The configurations all use the same \Sh{afl-clang-fast} with the grammar mutator as described above but with different scheduling strategies:\footnote{
  \url{https://aflplus.plus/docs/fuzzing_in_depth/}
}
\begin{description}
\item[$A_1$] uses the default exponential-power schedule \Sh{fast}~\cite{aflfast};
\item[$A_2$] uses the exploration-based constant-power schedule \Sh{explore};
\item[$A_3$] uses the quadratic-power schedule \Sh{quad}.
\end{description}

Consistently with \shf's setup, each experiment with \AFL ran one of its configuration for 12 hours on \mksh and was repeated three times.  We report the number of unique sanitizer warnings raised during the experiment, as well as the coverage of the shell interpreter.

\subsubsection{Crashes}
Another important consequence of the fact that \AFL may introduce arbitrary mutations -- as opposed to the constrained generation performed by \shf -- is that running a shell script produced by \AFL may incur a \emph{destructive} crash of the environment running the shell interpreter.  Unlike a ``benign'' crash, which simply results in an early termination of the interpreter process, a destructive crash compromises the environment to the extent that the testing process cannot continue -- for instance, by deleting the whole file system including the interpreter binaries.

 When a destructive crash occurs during an experiment, we collect any information about the experiment that survived the crash, we reset the environment, and then restart the experiment using any remaining time in the 12-hour slot.  Since each destructive crash requires a manual intervention, such crashes can severely worsen the overall efficiency of testing; even though we did not measure the time taken by such manual restarts, and report the results as if manual resets where instantaneous, we stress that they can be a significant hindrance in practice.

 We took some basic measures to reduce the impact of destructive crashes; in particular, we flagged as read-only all parts of the file system except the testing directory (where the scripts to be executed are stored).  Even though one could try to implement more sophisticated mitigation strategies, destructive crashes are always a possibility lest one precisely constraints the kinds of mutations that a generator can introduce (as we did with \shf's generators).

\begin{table}[!bth]
  \setlength{\tabcolsep}{3.5pt}
  \centering
  \begin{tabular}{lc rrrr rrrr rrrr}
    \toprule
    && \multicolumn{1}{c}{\textsc{sf}}
    & \multicolumn{1}{c}{$A_1$}
    & \multicolumn{1}{c}{$A_2$}
    & \multicolumn{1}{c}{$A_3$}
    & \multicolumn{1}{c}{\textsc{sf}}
    & \multicolumn{1}{c}{$A_1$}
    & \multicolumn{1}{c}{$A_2$}
    & \multicolumn{1}{c}{$A_3$}
    & \multicolumn{1}{c}{\textsc{sf}}
    & \multicolumn{1}{c}{$A_1$}
    & \multicolumn{1}{c}{$A_2$}
    & \multicolumn{1}{c}{$A_3$}
    \\
    \cmidrule(lr){3-14}
    &
    & \multicolumn{4}{c}{\textsc{maximum}}
    & \multicolumn{4}{c}{\textsc{average}}
    & \multicolumn{4}{c}{\textsc{minimum}}
    \\
    \cmidrule(lr){3-6}
    \cmidrule(lr){7-10}
    \cmidrule(lr){11-14}
    
    \textsl{destructive} & \#
    & 0 & 0 & 8 & 4
    & 0 & 0 & 3 & 1
    & 0 & 0 & 1 & 1
    \\
    \textsl{crashes} & \#
    & 155 & 238 & 200 & 156
    & 119 & 136 & 81 & 91
    & 92 & 83 & 0 & 4
    \\
    \textsl{coverage} & \%
    & 38 & 44 & 43 & 40
    & 38 & 42 & 37 & 37
    & 37 & 41 & 29 & 33
    \\[1mm]
    &
    & \multicolumn{4}{c}{\textsc{total}}
    & \multicolumn{4}{c}{\textsc{average}}
    & \multicolumn{4}{c}{\textsc{minimum}}
    \\
    \cmidrule(lr){3-6}
    \cmidrule(lr){7-10}
    \cmidrule(lr){11-14}
    
    \multirow{3}{*}{\textsl{confirmed}} & \textsl{all}
    & 7 & 4 & 4 & 4
    & 7 & 4 & 3 & 2
    & 7 & 4 & 2 & 0
    \\
    & \textsl{logic}
    & 5 & -- & -- & --
    & 5 & -- & -- & --
    & 5 & -- & -- & --
    \\
    & \textsl{memory}
    & 2 & 4 & 4 & 4
    & 2 & 4 & 3 & 2
    & 2 & 4 & 2 & 0
    
    \\
    \bottomrule
  \end{tabular}

  \caption{Comparison between \shf (\textsc{sf}) and three \AFL configurations ($A_1$, $A_2$, $A_3$). The table compares the number of \textsl{destructive} environment crashes, the number of ``benign'' interpreter crashes (\textsl{crashes}), the branch coverage of the shell interpreter, and the number of unique \textsl{confirmed} bugs (\textsl{all} categories, as well as the breakdown into \textsl{logic} and \textsl{memory} errors).  Each group of columns reports the \textsc{maximum}, \textsc{average} (mean), and \textsc{minimum} values per repeated run, or \textsc{total} in all runs.
  }
    \label{tab:comparison}
 \end{table}

 \subsubsection{Comparison Results}

 \autoref{tab:comparison}
 displays the main results of the comparison between \shf and the three \AFL configurations $A_1, A_2, A_3$.  The numbers for \shf refer to the same experiments described in greater detail in \autoref{sec:results:rq1} and \autoref{sec:results:rq2}.
 
 In terms of overall detected errors, \shf found 7 unique confirmed errors (5 logic errors and 2 memory errors), whereas \AFL found 4 (all memory errors).  As expected, \shf's main strength is not in detecting memory errors; however, it makes up for it with its unique capability of detecting logic errors.

 Whereas ``benign'' crashes occur with similar frequency both with \shf and with \AFL, destructive crashes only occur with \AFL.  Actually, \AFL configuration $A_1$ did not incur any destructive crashes in our experiments; however, there is no guarantee that it is immune from such crashes if one tried longer, or just repeated, testing sessions.\footnote{ We also noticed that even the ``benign'' crashes produced during experiments with $A_1$ usually had a bigger impact on the integrity of the environment than those occurred with \shf -- for example, by changing \AFL configuration files or environment variables.  }
 In contrast, a valuable advantage of \shf is that it was constrained to avoid destructive crashes by design -- which significantly helps the consistency and repeatability of a testing campaign.

 \begin{figure}[!hbt]
   \centering
   \includegraphics[width=12cm]{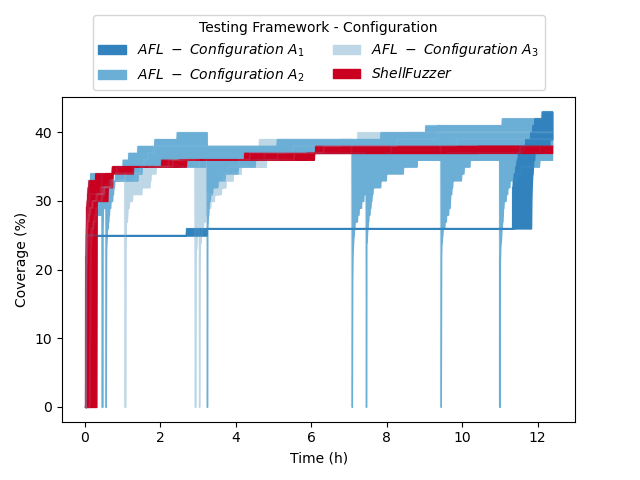}
   \caption{Coverage achieved by \shf and by \AFL's configurations $A_1, A_2, A_3$.  Each colored ribbon covers the range of coverage values reached in any experiment with that tool over the 12-hour time limit.}
\label{fig:coverage}
\end{figure}
 
 In terms of coverage, \shf trails behind \AFL by 2--6 percentage points (depending on the configuration).  Given that achieving a high coverage is not a main target of \shf's design, the gap is arguably to be expected and not tremendously consequential.  Furthermore, \shf's generation constraints (that avoid destructive crashes and enable detecting logic errors) inevitably somewhat reduce the reach of its exploration of the shell interpreter under test.

\autoref{fig:coverage} illustrates how the coverage evolves over time in each experiment.  \shf, as well as \AFL in configurations $A_2$ and $A_3$, reach a level of coverage close to the run's maximum within the first hour; then, coverage keeps increasing but much more slowly.  An important difference, however, is that \shf's coverage is always monotonically increasing; in contrast, it periodically resets to zero in \AFL's experiments whenever a destructive crash occurs -- given that the experiment has to restart from scratch.  \AFL's configuration $A_1$ is different still: while it does not have destructive crashes, its coverage grows kind of slowly and is well below \shf's up until the last hour of experiments, when it grows again prominently and even ``overtakes'' \shf in some repetitions.  We could not find a simple explanation for this peculiar behavior of \AFL, which is however a combined result of its exploration heuristics, the grammar that provides the testing seeds, and some random fluctuations.

Compared to \AFL, \shf demonstrates some advantages in terms of robustness/repeatability.  \autoref{tab:comparison} shows that the same unique confirmed bugs were revealed in all repeated runs of \shf; in contrast, only \AFL's configuration $A_1$ had a similar repeatability, whereas some repetitions of the experiments with configurations $A_2$ and $A_3$ failed to reveal some of the bugs.  On the same note, destructive crashes are also detrimental to robustness, as suggested by \autoref{fig:coverage}.

 \finding{
   In our experiments, \AFL usually achieved higher coverage than \shf (2--6\% higher), and revealed 2 more memory errors.  In contrast, \shf revealed 3 more errors overall (including 5 logic errors), and behaved more robustly (no destructive crashes, bugs detected in all repeated runs).
 }

\begin{table}[!bt]
  \centering
  \begin{tabular}{c *{3}{rrr}}
    \toprule
    \multicolumn{1}{r}{\textsc{generator}}
    & \multicolumn{3}{c}{\genV}
    & \multicolumn{3}{c}{\genI}
    & \multicolumn{3}{c}{\genM}
    \\
    \cmidrule(lr){2-4}
    \cmidrule(lr){5-7}
    \cmidrule(lr){8-10}
    
    \multicolumn{1}{r}{}
    & \multicolumn{1}{c}{\textsc{s}}
    & \multicolumn{1}{c}{\textsc{m}}
    & \multicolumn{1}{c}{\textsc{l}}
    & \multicolumn{1}{c}{\textsc{s}}
    & \multicolumn{1}{c}{\textsc{m}}
    & \multicolumn{1}{c}{\textsc{l}}
    & \multicolumn{1}{c}{\textsc{s}}
    & \multicolumn{1}{c}{\textsc{m}}
    & \multicolumn{1}{c}{\textsc{l}}
    \\
    \midrule
    \textsl{size}
    & 10 & 500 & 2500
    & -- & -- & --
    & 50 & 50 & 250 
    \\
    \textsl{procs}
    & 10 & 100 & 1000
    & 10 & 100 & 1000
    & 10 & 1000 & 1000
    \\
    \midrule
    \textsc{coverage}
    & 29\% & 29\% & 29\%
    & 34\% & 35\% & 35\%
    & 25\% & 27\% & 25\%
    \\
    \bottomrule
  \end{tabular}
  \caption{Parameter tuning of \shf. For each \textsc{generator}, we considered three combination of values \textsc{s}mall, \textsc{m}edium, and \textsc{l}arge for its parameters of script \textsl{size} (in lines of code) and number \textsc{procs} of parallel shell processes. The bottom line shows the resulting maximum \textsc{coverage} achieved with that choice of parameters.}
  \label{tab:parameters-range}
\end{table}

\subsection{RQ4: Parameter Tuning}
\label{sec:results:rq4}

In all the experiments described so far, \shf ran with the configuration described in \autoref{tab:parameters-default}.  Now, we discuss the experiments we performed to determine this standard configuration.

First of all, we tried to determine values for \autoref{tab:params_desc}'s parameters \textsl{size} (the approximate average size of the produced scripts, in lines of code) and \textsl{procs} (the number of shell processes running scripts in parallel during testing).  Instead of an exhaustive exploration of the full parameter space -- which would be unfeasible -- we determined, for each generator, three combinations of values for these two parameters roughly corresponding to ``small'', ``medium'', and ``large'' parameter values.  \autoref{tab:parameters-range} shows the nine parameter pairs (three for each generator) that we used in these tuning experiments.\footnote{ Remember that parameter \textsl{size} is immaterial for generator $I$, whose grammar determines the average size of the produced scripts.  }

Then, we ran each generator for 6 hours with each choice of parameter values, and measured the reached coverage.  \autoref{tab:parameters-range}'s bottom row shows the maximum coverage of each configuration's run.  We then selected each generator's parameters that achieved the best coverage; if two configurations achieved the same coverage, we preferred the one with higher throughput (measured as the number of scripts generated and executed per time unit).  In all, this led to choosing ``medium'' parameters for \genV and \genM, and ``large'' parameters for \genI.  We remark that \shf's behavior remained fairly consistent -- with only small changes in coverage and other measures -- even with significant changes to the parameter values.  Thus, the selected parameter values are a reasonable choice but they are by no means ``magic values''.

  \begin{table}[!bt]
    \setlength{\tabcolsep}{3pt}
    \small
    \centering
  \begin{tabular}{c l rr rr rr}
    \toprule
    & & \multicolumn{6}{c}{\textsc{generator time} [h]}
    \\
    \cmidrule(lr){3-8}
    & 
    & \multicolumn{2}{c}{$\ 4\genV + 4\genI + 4\genM\ $}
    & \multicolumn{2}{c}{$\ 4\genV + 6\genI + 2\genM\ $}
    & \multicolumn{2}{c}{$\ 5\genV + 5\genI + 2\genM\ $}
    \\
    \multicolumn{2}{c}{\textsc{category}}
    & \multicolumn{1}{c}{\textsc{sum}} & \multicolumn{1}{c}{\textsc{avg}}
    & \multicolumn{1}{c}{\textsc{sum}} & \multicolumn{1}{c}{\textsc{avg}}
    & \multicolumn{1}{c}{\textsc{sum}} & \multicolumn{1}{c}{\textsc{avg}}
    \\
    \cmidrule(){1-2}
    \cmidrule(l){3-4}
    \cmidrule(l){5-6}
    \cmidrule(l){7-8}
    
    \multirow{4}{*}{\textsl{all}}
    &
        \textsl{alerts} & 153205 & 51068.3
                     & 203460 & 67820.0
                     & 158230 & 52743.3
        \\
        & \textsl{unique}  & 12 & 11.3
                & 11 & 10.3
                & 11 & 11.0
        \\
        & \textsl{true positives} & 8 & 8.0
                      & 8 & 8.0
                      & 8 & 8.0
        \\
        & \textsl{confirmed}  & 7 & 7.0
                   & 7 & 7.0
                   & 7 & 7.0
    \\
    \cmidrule(){1-2}
    \cmidrule(l){3-4}
    \cmidrule(l){5-6}
    \cmidrule(l){7-8}
    
    \multirow{4}{*}{\textsl{logic}}
    &
        \textsl{alerts} & 152919 & 50973.0
                     & 203104 & 67701.3 
                     & 157820 & 52606.7
        \\
        & \textsl{unique} & 10 & 9.3
                & 9 & 8.3
                & 9 & 9.0
        \\
        & \textsl{true positive}  & 6 & 6.0
                       & 6 & 6.0
                       & 6 & 6.0
        \\
        & \textsl{confirmed}  & 5 & 5.0 
                   & 5 & 5.0 
                   & 5 & 5.0 
    \\
    \cmidrule(){1-2}
    \cmidrule(l){3-4}
    \cmidrule(l){5-6}
    \cmidrule(l){7-8}
    
    \multirow{4}{*}{\textsl{memory}}
    &
        \textsl{alerts}  & 286 & 95.3
                    & 356 & 118.7
                    & 410 & 136.7
        \\
        & \textsl{unique} & 2 & 2.0
               & 2 & 2.0
               & 2 & 2.0  
        \\
        & \textsl{true positive} & 2 & 2.0
                      & 2 & 2.0 
                      & 2 & 2.0 
        \\
        & \textsl{confirmed} & 2 & 2.0 
                  & 2 & 2.0 
                  & 2 & 2.0 
    \\
    \cmidrule(){1-2}
    \cmidrule(l){3-4}
    \cmidrule(l){5-6}
    \cmidrule(l){7-8}
    
    \multicolumn{2}{c}{\textsc{coverage}}
    &  \multicolumn{2}{c}{37.7 \%}
    &  \multicolumn{2}{c}{37.9 \%}
    &  \multicolumn{2}{c}{37.7 \%}
    \\
    \bottomrule
	\end{tabular}
   \caption{Effectiveness of \shf with different \textsc{time} allocated to each \textsc{generator}.
     $x\genV + y\genI + z\genM$ denotes experiments where \genV ran for $x$ hours, \genI for $y$ hours, and \genM for $z$ hours, such that $x + y + z = 12$.  For each such time partitioning, the table reports the total number \textsc{sum} over three repeated runs, and average \textsc{avg} per repeated run of \textsl{alerts} raised by \shf, \textsl{unique} alerts, \textsl{true positive} bugs, and bugs \textsl{confirmed} by the shell maintainers.  The three row groups report data for \textsl{all} alerts, as well as the breakdown into \textsl{logic} and \textsl{memory} alerts (identified as described in \autoref{sec:methodology}).  Finally, the bottom row reports the maximum branch \textsc{coverage} reached during any run.
   }
 \label{tab:effectiveness-time-share}
 \end{table}

Finally, we selected values for
\autoref{tab:params_desc}'s parameter \textsl{time}:
the time allotted to each generator within
a standard 12-hour session.
We considered three different partitioning of time:
\begin{description}
\item[$4\genV + 4\genI + 4\genM$:] each generator gets the same amount of time (4 hours);
\item[$4\genV + 6\genI + 2\genM$:] generator \genI gets the longest time slot (6 hours), followed by \genV (4 hours),
  and then \genM (2 hours);
\item[$5\genV + 5\genI + 2\genM$:] generators \genV and \genI get the bulk of testing time (5 hours each), \genM gets the remaining time (2 hours).
\end{description}
Given that it's clear that running all generators is beneficial, we did not consider configurations where a generator does not run at all; besides, we usually give less time to generator \genM, since it is more specialized, and hence it is unlikely to benefit from much longer running times.

We ran three repetitions of the experiments with each time configuration; \autoref{tab:effectiveness-time-share} shows the results of these experiments.  Once again, the results in terms of coverage and detected bugs are quite similar independent of the time partitioning that we adopt.  In the end, we selected configuration $4\genV + 6\genI + 2\genM$ since it achieves a slightly higher coverage and precision ($\textsl{true positives}/\textsl{unique}$) than the other configurations.  In fact, the central section of \autoref{tab:effectiveness-time-share}
(configuration $4\genV + 6\genI + 2\genM$) corresponds to the same experiments reported in \autoref{tab:summary_generators_v2} and discussed
in the previous parts of \shf's experimental evaluation.

\finding{
  Choosing different values for its parameters (script size, parallel processes, and time for each generator) did not significantly affect \shf's effectiveness.
}

\subsection{Threats to Validity}
\label{sec:threats}

We review threats to the validity of the experiments described in this section, as well as how we mitigated them.

\paragraph{Construct validity}
Given the nature of our experiments, construct validity mainly pertains the choice of metrics used to study \shf's capabilities -- which are all fundamental, widely used measures.  To study \shf's effectiveness as a bug-finding tool, we focused on the key metric of (confirmed) true positive detected errors.  Measures related to alert uniqueness and script minimization are also customary to assess effectiveness and how a tool's output is accessible for users.  Finally, we also considered code coverage; while it is only a weak proxy for the more important metric of ``found bugs'', it remains widely used as it gives a rough idea of the thoroughness of the input space search.

\paragraph{Internal validity}
Internal validity depends on the presence of possible confounding factors.  To account for the randomness of the script-generation process, we repeated each experiment three times, and reported averages and totals over all runs.  In our main experiments, we empirically chose the parameter values used to configure \shf  based on the results of separate tuning experiments (\autoref{sec:results:rq4}).  It is important to notice that in the tuning experiments, as well as in the repeated runs of the main experiments, we observed remarkably \emph{consistent} results from run to run.  Hence, while it is possible that configuring \shf very differently would yield different results, we are reasonably confident that our paper's findings are sufficiently representative and informative.

Determining which of the (unique) alerts raised during \shf's experiments are true positives involved some manual analysis.  We took standard measures -- such as performing multiple checks, trying to reproduce issues on different shell interpreters, and consulting different sources of documentation of expected behavior -- to reduce the risk that our judgment was biased or unreliable.  Ultimately, we tried to get every alert that we classified as true positive confirmed by the shell's maintainers.  Even though deciding whether a certain interpreter behavior is acceptable or not may involve criteria that go beyond strict adherence to the POSIX standard, it remains that most of the issues we reported to maintainers were confirmed (and, in many cases, were addressed by changes in the code or documentation of the interpreter); this is reassuring evidence that the most significant threats to internal validity were adequately mitigated.

In order to make the comparison between \shf and \AFL fair,
we configured \AFL with the same grammar mutator used by \shf's implementation, and we adopted the same experimental conditions (three repetitions of 12-hour testing sessions).  Since \AFL is a highly configurable tool, we performed experiments with three different exploration heuristics and reported all results.  Since \AFL may occasionally generate scripts that destroy the testing environment, we also took measures to limit their impact, and discounted from the overall testing time the interruptions needed to (manually) reset the environment and continue testing.  Despite this process, \AFL and \shf remain tools with distinct strengths and weaknesses: the goal of our comparison is demonstrating what each tool does best, and how they differ.

\paragraph{External validity}

External validity concerns the generalizability of the research findings.  In our experiments, we primarily targeted the \mksh shell, because it is a widely used POSIX-compliant shell with a somewhat shorter development history compared to other de facto standard shells.  Nevertheless, \shf is applicable to test any other shell interpreter -- in fact, we found that two \mksh memory errors that we discovered in our experiments also affect Bash.

We focused on a subset of the POSIX standard shell language because it underlies a large number of widely used shell interpreters.  By changing the grammars used by \shf's generators, one could adapt \shf to work with different kinds of shell languages (for example, the Windows PowerShell), or even other interpreted languages (for example, Python or JavaScript).  Of course, effectively supporting very different languages may require to adjust other features of \shf's design.  A crucial feature is the automated generation of oracles based on the expected validity or invalidity of a generated program (\autoref{sec:methodology}); this may be trickier to implement when targeting languages with stricter standards, single implementations, or no legacy features.

\section{Conclusions and Future Work}
\label{sec:conclusion}
\label{sec:discussion}
We presented \shf: an automated testing technique to find bugs of different kinds in shell interpreters.  \shf's design addresses several of the challenges of testing a shell interpreter -- some shared with testing interpreters or compilers for other languages, others specific to the shell language.  To produce a broad variety of \emph{structured} inputs -- i.e., POSIX-compliant shell scripts -- \shf combines different grammar-based generators with some carefully chosen random mutation.  Each generator produces scripts with homogeneous characteristics, which makes it possible to also automatically produce an oracle that states the expected overall behavior (error/no error) of executing the script, whereas the combination of three generators increases the overall diversity.  Another key design choice were the mutations and the fragment of the POSIX grammar that each generator uses: we chose them to minimize the chance that a generated script incurs destructive behavior -- irreversibly damaging the testing environment.  Our experiments demonstrate that \shf is effective at finding bugs of various kinds in widely used shell interpreters (i.e., \mksh and Bash).  Even though none of the bugs we found are likely to have a major impact on the security of using Unix shells, finding any kind of developer-confirmed misbehavior is non-trivial for software with such a long usage and development history.

The comparison between \shf and \AFL indicated that the two approaches have largely complementary features.  Of course, \AFL is a much more mature and robust tool than \shf, and remains remarkably effective as a general-purpose fuzzer~\cite{fuzzing_survey}.  \shf is specifically geared towards detecting \emph{logic} bugs -- for example affecting the behavior of the parser or other interpreter features.  In contrast, \AFL is usually more effective at achieving a high coverage, which can be a quicker way of discovering ``fundamental'' bugs such as memory-related errors.  The fact that \shf managed to avoid destructive crashes in all the experiments is a strength of our approach, since it makes for a more automated, uninterrupted testing process.

\subsection*{Future Work}

A straightforward way of improving \shf's effectiveness is using additional implicit oracles -- such as sanitizers for concurrency issues or security properties.  This does not require, in principle, to modify anything in how \shf generated shell scripts; however, certain kinds of bugs may only be triggered by using certain kinds of shell features.

While different shell interpreters have different levels of robustness and maturity, there are no a priori reasons that prevent us from extending the experiments with \shf to other shells, including those with non POSIX-comp\-at\-i\-ble extensions.  Doing so would require modifying or extending the grammars used by \shf's generators, as well as their mutation operators.

A key challenge would be balancing out the expressiveness of the generation process and retaining some control over it, so as to avoid destructive behavior and to automatically build oracles.  In \shf, we explicitly designed the generators with this balance in mind.  In future work, one may try to generate the grammars and mutations themselves in some sort of automated way, for example by relying on a formal semantics of the shell~\cite{shell_semantics} or by mining them from examples~\cite{CLIfuzzer,no-samples}.

\section*{Declaration of competing interest}
The authors declare that they have no known competing financial interests or personal relationships that could have affected the work reported in this paper.

\section*{Data Availability}
The prototype implementation of \shf is available in a public repository.

\section*{Acknowledgments}
The authors gratefully acknowledge the financial support of the Swiss National Science Foundation for the work (SNF grant 200020-188613).


%
%

\bibliographystyle{abbrv}

%
%

\end{document}